**Engineering magnetic anisotropy and ferromagnetism in topological Kagome metal $GdV_6Sn_6$ via Nd substitution**

Santosh Karki Chhetri[1], M.M. Sharma[1*], Jian Wang[2], Dinesh Upreti[1], Gokul Acharya[1], Md Rafique Un Nabi[1], Jin Hu[1,3,4,5#]

[1]*Department of Physics, University of Arkansas, Fayetteville, Arkansas 72701, USA.*

[2]*Department of Chemistry and Biochemistry, Wichita State University, Wichita, Kansas 67260, USA.*

[3]*Arkansas Materials Institute, Institute for Nanoscience and Engineering, University of Arkansas, Fayetteville, Arkansas 72701, USA.*

[4]*MonArk NSF Quantum Foundry, University of Arkansas, Fayetteville, Arkansas, 72701, USA.*

[5]*Smart Ferroic Materials Center, University of Arkansas, Fayetteville, Arkansas, 72701, USA.*

**Abstract:**

Kagome metals with the formula $RM_6X_6$ ($R$ = rare-earth, $M$ = 3d transition metal, and $X$ = Sn/Ge) provide a rich platform for exploring magnetic and electronic phenomena, with tunable properties enabled by the combination of rare-earth elements and transition metals. In this study, we report the structural, electrical and magnetic properties of Kagome metal $(Nd_xGd_{1-x})V_6Sn_6$. We demonstrate that substituting lighter Nd atoms at the Gd site tunes the complex magnetic ground state of $GdV_6Sn_6$ into a ferromagnetic-like one. Moreover, the isotropic magnetization of $GdV_6Sn_6$ becomes anisotropic, with the *c*-axis emerging as the easy axis. Transport measurements reveal a strong coupling between magnetism and electronic properties, with negative magnetoresistance observed at low magnetic fields in all compositions. In addition, a fourfold anisotropy component tends to emerge in end compounds at higher magnetic fields. These findings highlight the role of rare-earth substitution in tuning magnetic anisotropy and magneto-transport behaviour in $RM_6X_6$ compounds, featuring a non-magnetic Kagome layer, with potential implications for spintronic and topological applications.

[*]ms207@uark.edu

[#]jinhu@uark.edu

## Introduction:

The interplay of topology, electron correlations, and magnetism drives the emergence of exotic quantum phases in solid-state systems [1–4]. The Kagome lattice, formed by corner-sharing triangles, naturally integrates these ingredients, providing a versatile platform to investigate quantum phenomena such as unconventional superconductivity [4,5], charge and spin density waves [6,7], quantum spin liquids [8,9], frustrated magnetic interactions [10], and high carrier mobility [11,12]. Its intricate electronic band structure features Dirac and Weyl points [13–17], nodal lines [18], van Hove singularities [19,20], and topological flat bands [21,22], further enhances the opportunities for discovering new quantum states. Recently, materials that combine bulk magnetism with non-trivial electronic structures have attracted significant interest, driven by predictions of phenomena such as the quantum anomalous Hall effect [23,24] and topological magneto-electric effect [25,26]. Therefore, Kagome magnets have emerged as a promising platform that enables exploration of the intricate interplay between long-range magnetic order and non-trivial topological electronic states [16,21,27–30].

$RM_6X_6$ ($R$ = rare-earth, $M$ = 3d transition metal, and $X$ = Sn/Ge), often termed as *R166,* is a member of Kagome magnets. As shown in Fig. 1(a), these materials crystallize in the $HfFe_6Ge_6$-type (P6/mmm) crystal structure, where the transition metal atoms form Kagome planar layers. The rare-earth atoms occupy interlayer sites between adjacent Kagome sheets along the $c$-axis [31,32]. The *R166* family exhibits a strong interaction between the $d$-electrons of the Kagome layer and the localized $f$-electrons of the rare earth sublattice [32]. Most of the studied *R166* materials are based on Kagome lattice of Mn, exhibiting complex magnetism due to the presence of multiple magnetic interactions emerging from Mn-Mn, $R$-Mn and $R$-$R$ couplings [33,34]. The Mn-Mn interaction within the Kagome layer exhibits ferromagnetic alignment, while its interaction with the rare-earth moments ($R$-Mn coupling) leads to ferrimagnetic ordering. The magnetic anisotropy in these systems is primarily determined by the choice of the rare-earth sublattice [35]. Furthermore, the incorporation of lighter rare-earth elements (Nd and Sm) elongates the unit cell along the *c*-axis and distorts the Kagome lattice [31]. Also, rare-earth engineering has been an effective strategy to tune the topological electronic states in Mn-based *R*166 compounds [36,37].

In addition to $RMn_6Sn_6$, the Kagome lattice formed by non-magnetic V atoms ($RV_6Sn_6$) provides a less complicated model, as its magnetism only arises from interlayer and intralayer *R*-*R* magnetic interactions [38]. Though less explored, the reported $RV_6Sn_6$ compounds exhibit many interesting properties, such as tunable Dirac states and van Hove singularity in $GdV_6Sn_6$ [39], finite spin Berry curvature in (Tb, Ho, Sc)$V_6Sn_6$ [40], and a triangular Kondo lattice in $YbV_6Sn_6$ [41]. Particularly, an interesting feature of $RV_6Sn_6$ is the presence of a flat electronic band close to the Fermi level. The energy position of the flat band is determined by the hybridization between the *d*-orbitals of V and the *p*-orbitals of Sn [42]. This hybridization, in turn, is influenced by the chemical pressure related to the choice of the rare-earth element. Additionally, similar to $RMn_6Sn_6$, in which the Mn moment orientation is controlled by the choice of rare earth, the choice of *R* also determines the magnetic anisotropy in $RV_6Sn_6$, though magnetism in these V-based compounds only originates from the rare earth [38].

Providing that rare earth engineering in $RM_6X_6$ is highly effective in tuning the structural, electronic and magnetic properties, in this work we investigated the impact of Nd substitution for Gd in $GdV_6Sn_6$, which eventually leads to $NdV_6Sn_6$, a new member of the $RV_6Sn_6$ family. We found multiple rare earth and Sn sites when Nd content is high. Magnetization measurements reveal that Nd substitution leads to a change in magnetic anisotropy and stabilizes a ferromagnetic-like state. Magnetotransport displays crossover from negative magnetoresistance (MR) to positive MR, characterized by two- and four-fold MR anisotropy and reflecting the roles of magnetism and Fermi surface. Our findings highlight the efficiency of rare-earth engineering in $RV_6Sn_6$ to control magnetic and transport properties, opening avenues for designing materials with tunable and emergent magnetic functionalities.

**Experiment:**

Single crystals of $Nd_xGd_{1-x}V_6Sn_6$ were synthesized using a Sn-flux method. High-purity powders of Nd, Gd, V and Sn were vacuum sealed in quartz ampoules. The quartz ampoule was then heated to 1150 ºC at a heating rate of 50 ºC/h, kept for 48 hours, slowly cooled to 730 ºC at a rate of 2 ºC/h, and centrifuged to separate crystals from Sn flux. The crystal structure and stoichiometry of the synthesized compounds was verified by X-ray diffraction (XRD) and energy dispersive X-ray analysis (EDS), respectively. Magnetic measurements were performed using a Magnetic Property Measurement System (MPMS3 SQUID), and the magneto-transport measurements were performed on a Physical Property Measurement System (PPMS).

**Results and Discussion:**

Composition analysis for $Nd_xGd_{1-x}V_6Sn_6$ single crystals by EDS reveals actual compositions that are close to the nominal ones, as summarized in Table 1. There are very few reports on $RM_6X_6$ materials containing the lighter rare-earth elements positioned to the left of Gd in the periodic table [31,42]. Previous reports indicate that incorporating lighter rare-earth elements induces structural distortion. For instance, in $SmV_6Sn_6$, Sm atoms occupy two distinct atomic positions [42]. This distortion becomes more pronounced in Mn-based $SmMn_6Sn_6$ and $NdMn_6Sn_6$, where the space group changes from *P6/mmm* to *Immm*, resulting in a distorted Kagome lattice of Mn atoms [31]. To investigate the potential structural modification induced by Nd substitution in $GdV_6Sn_6$, the crystal structures of the obtained $Nd_xGd_{1-x}V_6Sn_6$ crystals were determined by single-crystal XRD (SXRD) refinement. Notably, no changes in the space group are observed, even in the fully substituted $NdV_6Sn_6$ compound, indicating that the system retains the *P6/mmm* space group symmetry against Nd substitution in V-based *R166* Kagome materials. This observation is in sharp contrast to the previously reported distortions in other light rare-earth based *R166* compounds mentioned above. Nevertheless, high-Nd (above x = 0.5) samples display multiple Nd and Sn sites (Fig. 1(b)) which effectively form disorders, as summarized in Table 1. However, similar site modifications may already begin at lower Nd concentrations, but despite repeated synthesis attempts, high-quality single crystals in the intermediate composition range ($0.05 < x < 0.5$) suitable for structural and transport measurements could not be obtained. A comparable structural feature has been reported in $SmV_6Sn_6$, where two distinct Sm sites result in a combination of two $HfFe_6Ge_6$-type structures translated by $c/2$ along the $c$-axis. This structural modulation has been suggested to suppress long-range magnetic order in $SmV_6Sn_6$. In $Nd_xGd_{1-x}V_6Sn_6$ with a high amount of Nd, however, the rare-earth atomic site undergoes only a tiny translation along $c$, and the long-range magnetic order is still present, as will be shown later.

Figures 2(a) and 2(b) depict the temperature-dependent in-plane ($\chi_{ab}$) and out-of-plane ($\chi_c$) magnetic susceptibility for $Nd_xGd_{1-x}V_6Sn_6$ single crystals, measured with a magnetic field of 0.05 T applied within the *ab*-plane and along the *c*-axis, respectively. For $GdV_6Sn_6$, the magnetic susceptibility shows a transition at 5 K, consistent with the earlier studies [43,44]. Under this field, irreversibility between zero-field cooled (*ZFC*) and field-cooled (*FC*) measurements is negligible. However, irreversibility becomes notable in low field measurements (Figs. 2a and 2b, inset), which agrees with the previous observations [43] and suggests soft ferromagnetism that is consistent with isothermal magnetization (see below). It is worth noting that the reduction of $\chi_{ab}$ below ordering temperature for both *ZFC* and *FC* appears in line with antiferromagnetism, so the magnetic ground state in $GdV_6Sn_6$ may involve a more complicated moment configuration. Nevertheless, the greater $\chi_{ab}$ suggests *ab*-plane as the magnetic easy-plane. Nd substitution reduces transition temperature and susceptibility, while induces pronounced irreversibility. Comparing $\chi_{ab}$ and $\chi_c$, a change from easy-plane (*ab*-plane) to easy-axis (*c*-axis) is likely to occur with Nd substitution, which is further supported by the isothermal magnetization measurements. As shown in Fig. 2(c), the in-plane and out-of-plane magnetization for $GdV_6Sn_6$ essentially overlap at high field, displaying a polarization behavior at low field. Such nearly isotropic magnetization in $GdV_6Sn_6$ can be understood in terms of vanishing spin-orbit coupling (*SOC*) and quenched orbital angular momentum of *4f* orbital of $Gd^{3+}$ ions [45]. Substituting $Gd^{3+}$ for $Nd^{3+}$ with unquenched *4f* orbital angular momentum gradually introduces magnetic anisotropy, with the saturation magnetization becoming more pronounced when *H*//*c* and signaturing an easy-axis along *c*. Such modulation of magnetic anisotropy is in line with the earlier reports on the role of rare earth elements in affecting the easy axes in other *R166* materials [38,46]. To further investigate the nature of magnetic ordering in $NdV_6Sn_6$, Arrott plot analysis ($M^2$ vs $H/M$) was performed at low temperatures, as shown in Fig. S1(a) in supplementary materials [47]. The plots exhibit noticeable curvature at low magnetic fields, indicating deviation from ideal mean-field ferromagnetic behaviour and suggesting the absence of a well-defined spontaneous magnetization. Under higher magnetic fields, the curves become nearly linear and approximately parallel as shown in Fig, S1(b) [47], indicating the presence of a ferromagnetic-like state in $NdV_6Sn_6$, which is stabilized primarily by the applied magnetic field rather than arising from the spontaneous ferromagnetic ordering.

Temperature-dependent resistivity measurements for $Nd_xGd_{1-x}V_6Sn_6$ single crystals are shown in Fig. S2(a)–(e) in Supplementary Materials [47], with the normalized resistivity curves for all samples presented in Fig. 3. A clear low temperature resistivity kink at the AFM transition temperature is observed for the Gd-rich compounds ($x = 0$ and 0.05), consistent with the previous studies on pristine $GdV_6Sn_6$ [43,48,49]. Interestingly, this resistivity kink is significantly suppressed in FM samples with high Nd content (Fig. 3, inset), which might be associated with the reduced magnetic ordering temperature and moments with Nd substitution. The impact of magnetism on transport is further investigated by magneto-transport measurements, as shown in Figs. 4(a)–(e). MR has been calculated by normalizing the field-dependent resistivity with zero field resistivity using the formula $MR = [\rho(B) - \rho(0)]/\rho(0)$, where $\rho(0)$ is resistivity at zero magnetic field. $GdV_6Sn_6$ exhibits a sharp resistivity drop at low fields that forms a MR peak at zero field. Such low-field negative MR agrees with the earlier reports [48,49]. The sharp resistivity drop coincides with the magnetization polarization, as shown in Fig. 4(f). At higher field when magnetization is fully saturated, resistivity displays an upturn and eventually crossover to positive MR at higher fields. Such field dependence implies that the suppression of magnetic scattering by spin polarization likely results in negative MR at low field, which competes with a positive MR component that becomes predominant at higher field. At higher temperatures, the low field negative MR in $GdV_6Sn_6$ persists above 20K, well above the magnetic transition temperature (~5 K), which further support the suppression of spin scattering as the origin for negative MR because spin scattering can remain significant above the magnetic ordering temperature, as seen in [50]. Such spin scattering scenario should be the most significant near the magnetic transition due to the strongest magnetic fluctuations. Indeed, the negative MR is the strongest near the magnetic transition. Nd substitution suppresses magnetic ordering temperature from 5 K in $GdV_6Sn_6$ to 2 K in $NdV_6Sn_6$, and coincidently, the negative MR is the strongest at 5 K in $GdV_6Sn_6$ (Fig. 4a) but at 2 K in $NdV_6Sn_6$ (Fig. 4e).

Next, we discuss the positive MR component. The negative MR component, which originates from the field-suppression of spin scattering, should saturate when spins are fully polarized. Therefore, the crossover to positive MR indicates the robust growth of the positive MR component with increasing magnetic field. As shown in Fig. 4, the nearly quadratic field dependence of the high field positive MR component suggests a classical orbital MR origin. Estimating from the MR crossover from negative to positive, the positive MR component reaches nearly 15% at 9 T and 2 K for Gd-rich samples. Within the framework of orbital MR, this MR magnitude is considerable and suggestive of sizable mobility. To evaluate mobility, we performed Hall effect measurements for $Nd_xGd_{1-x}V_6Sn_6$, as shown in Fig. 5(a–e). For all samples, Hall resistivity ($\rho_{yx}$) exhibits nonlinear field dependence below 100 K, indicating multi-band transport. Above 100 K, $\rho_{xy}$ becomes linear in field with a positive slope, suggesting hole-dominant transport. To quantitative analysis Hall effect, the two-band model has been used:

$$\rho_{xx} = \frac{1}{e} * \frac{(n_h\mu_h+n_e\mu_e)+\mu_e\mu_h*(n_h\mu_e+n_e\mu_h)*B^2}{(n_h\mu_h+n_e\mu_e)^2+\mu_e^2\mu_h^2*(n_h-n_e)^2*B^2} \quad (1)$$

$$\rho_{xy} = \frac{B}{e} * \frac{(n_h\mu_h^2-n_e\mu_e^2)+\mu_e^2\mu_h^2*(n_h-n_e)*B^2}{(n_h\mu_h+n_e\mu_e)^2+\mu_e^2\mu_h^2*(n_h-n_e)^2*B^2} \quad (2)$$

where $n_h$ and $n_e$ are the hole and electron densities, and $\mu_h$ and $\mu_e$ represent their respective mobilities. In principle, simultaneous fits for both $\rho_{xx}$ and $\rho_{xy}$ are necessary for reliable analysis [51]. However, this classical model assumes a positive and parabolic dependence of $\rho_{xx}$ without the consideration of magnetic scattering, which is not applicable for our low temperature magnetotransport here. Therefore, we focus on data at 100 K where MR is positive for the entire field range. The two-band model nicely reproduces both $\rho_{xx}$ and $\rho_{yx}$ (Fig. S3, Supplementary Materials) [47], yielding carrier densities and mobilities as shown in the upper panels of Fig. 5(f) and (g), respectively. The electron and hole densities are found to be in the order of $10^{21}$ $cm^{-3}$ and $10^{20}$ $cm^{-3}$ respectively for all samples, indicating that electron and hole carriers are not compensated at this temperature. The extracted electron and hole mobilities at 100 K are 157 $cm^2$/V-s and 300 $cm^2$/V-s respectively for $GdV_6Sn_6$, which drops by nearly 1/3 in $NdV_6Sn_6$. For non-magnetic simple metals, carrier mobility is expected to remarkably enhance with the significant reduction of phonon scattering from 100 K to 2 K. This scenario is also applicable for magnetic $Nd_xGd_{1-x}V_6Sn_6$ when magnetic scattering is substantially suppressed by spin polarization under field, which accounts for the sizable positive MR component at low temperatures and high fields. Above 100K where transport is dominated by holes, the linear $\rho_{yx}(H)$ allows for direct estimation with a single-band model, in which the slope of $\rho_{yx}(H)$, i.e., Hall coefficient $R_H$, is related to the carrier density $n$ as $R_H = \frac{1}{ne}$. From the obtained carrier density, mobility can be extracted via $\mu = \frac{1}{\rho_{xx} ne}$, as shown in the lower panels of Fig. 5(f) (200 K) and Fig. 5(g) (300 K). Both carrier density and mobility do not show very strong variation with increasing Nd concentration. This indicates that the isovalent Nd substitution might not significantly influence the electronic structure in the paramagnetic state. One important feature observed in the Hall measurements of $GdV_6Sn_6$ and $NdV_6Sn_6$ is that, despite the presence of strong magnetism, no discernible magnetic contribution to the Hall resistivity is detected. This indicates that finite magnetization alone might not be sufficient to generate a measurable anomalous Hall signal. Similar behavior has been observed in the Heusler compound $Mn_2CoGa$, where a significant magnetic moment coexists with strongly suppressed anomalous Hall conductivity due to negligible net Berry curvature in the electronic structure arising from compensating contributions from spin-up and spin-down electrons [52]. This analogy may be relevant in the present system, where cancellation of Berry-curvature contributions near the Fermi level could suppress the anomalous Hall response. In addition to intrinsic mechanisms, extrinsic mechanisms such as skew scattering may also lead to AHE. Additional studies on electronic structures of these V-based *R166* compounds and scattering mechanisms are needed for the clarification of this intriguing absence of an anomalous Hall effect.

Since Nd substitution in $GdV_6Sn_6$ mainly affects magnetism and low temperature magnetotransport, we examine the anisotropy of magnetotransport. Because $GdV_6Sn_6$ and $NdV_6Sn_6$ are characterized by distinct magnetic anisotropy, we focused on these two end compounds in this work. Figure 6(a) shows the field-dependent MR for $GdV_6Sn_6$ measured with various field orientations, with the schematic of measurement depicted in the inset. In the low field region below 1 T, the measured MR curves collapse into a single peak for all measured angles, indicating isotropic negative MR that agrees with the nearly isotropic magnetization for $GdV_6Sn_6$ (Fig. 2c). MR anisotropy starts to develop above 1 T and enhances with field. In contrast, MR anisotropy for $NdV_6Sn_6$ can be observed under a much lower field. As shown in Fig. 6(b), the low-field MR peak for $NdV_6Sn_6$ is the sharpest under the out-of-plane field ($\theta$ = 0) and broadens with the rotation of the field away. Such observation agrees with the significantly enhanced magnetic anisotropy and the out-of-plane magnetic easy axis in $NdV_6Sn_6$. To gain more insight into the anisotropic MR, in Figs. 6(c) and (d) we present the angular dependent MR (AMR) measured at various fixed magnetic fields for $GdV_6Sn_6$ and $NdV_6Sn_6$, respectively. Here AMR is defined as $[\rho(\theta)-\rho_{min}]/\rho_{min}$ where $\rho_{min}$ is the minimum resistivity during the rotation of magnetic field, which takes value of $\rho(\theta=90°)$ for $GdV_6Sn_6$ and $\rho(\theta=0°)$ for $NdV_6Sn_6$. For $GdV_6Sn_6$ (Figs. 6(c)), AMR at 1 T displays two-fold anisotropy that reaches a maximum at $\theta = 0°$ (out-of-plane field). However, the AMR maximum at 0° splits at higher fields, manifested into two peaks. Interestingly, a similar evolution of AMR anisotropy with field is also seen for $NdV_6Sn_6$ (Fig. 6(d)), but the low-field AMR maximum occurs at $\theta = 90°$ (in-plane field). Such a change coincides with the switching of the magnetic easy axis from the *ab*-plane in $GdV_6Sn_6$ to the *c*-axis in $NdV_6Sn_6$, implying magnetism plays a role in mediating magnetotransport at low field. To better visualize the field induced anisotropy change, contour plots showing the magnitude of AMR as a function of magnetic field and field orientation are shown in Figs. 6(g) and (h) for $GdV_6Sn_6$ and $NdV_6Sn_6$ respectively, from which a crossover from two-fold to four-fold is clearly visible, as marked by white dashed lines. To quantify such field-induced anisotropy change, we fitted AMR with an anisotropy equation [53]:

$$\mathrm{AMR} = C_1 + C_2 \cos 2(\theta + \delta_2) + C_4 \cos 4(\theta + \delta_4) \quad (3)$$

where $C_2$ and $C_4$ take positive values and represent the weights of the two- and four-fold components respectively, and $\delta_2$ and $\delta_4$ are the phase factors of the two- and four-fold components. As shown in Figs. 6(c) and (d), the above equation reproduces AMR for both $GdV_6Sn_6$ and $NdV_6Sn_6$ very well. The evolution from two-fold to four-fold anisotropy can be cleanly captured by the enhanced weight ratio of four- and two-fold $C_2/C_4$ with increasing magnetic field, as shown in the upper panels of Fig. 6(e) and (f). The phase factors also provide useful information. $\delta_2$ is essentially field-independent but differs by 90° between $GdV_6Sn_6$ and $NdV_6Sn_6$, representing the 90° shift of the AMR maximum between these two compounds. Nevertheless, $\delta_4$ takes similar values for these two compounds, indicating that the four-fold anisotropy in $GdV_6Sn_6$ and $NdV_6Sn_6$ might have the same mechanism.

As shown above (Fig. 4), $GdV_6Sn_6$ and $NdV_6Sn_6$ both exhibit crossovers from negative MR to positive MR with increasing magnetic field. Noticing that the four-fold AMR anisotropy gradually emerges with the development of the positive MR, hence the low-field two-fold and high-field four-fold AMR anisotropies should be attributed to the negative and positive MR components, respectively. As discussed above, the low-field negative MR component arises from the suppression of spin scattering by magnetic field, which should follow the anisotropy of magnetism. Indeed, the low-field two-fold anisotropy in $GdV_6Sn_6$ and $NdV_6Sn_6$ differs by 90°, which is consistent with their distinct magnetic easy axis as discussed above.

Next, we turn to the high-field four-fold anisotropy for the positive MR component. Four-fold AMR anisotropy has been observed in various materials with different geometries used for the rotation of magnetic fields. Several mechanisms have been proposed, such as magneto-crystalline anisotropy [54], spin-dependent scattering near antiphase boundaries [53], anisotropic Fermi surface [55–59] and modulation of the density of states (DOS) near the Fermi energy [60,61]. The first two mechanisms can be excluded. As discussed above, magnetism in these materials should lead to negative MR with two-fold anisotropy. Owing to the absence of a field-induced metamagnetic transition, magnetism alone is not expected to generate positive MR with four-fold anisotropy. Also, the anisotropic magnetism-induced effect should disappear at the magnetic fields that are higher than the field at which the moment becomes polarized. Due to this, we attribute the observed fourfold anisotropy to the anisotropic Fermi surface and modulation of DOS near the Fermi level. In the presence of a magnetic field, the charge carriers conduct cyclotron orbits along cross-sections of the Fermi surface normal to the field. For an isotropic Fermi surface, such orbital MR should not display any anisotropy, except for that from the Lorentz effect. Therefore, a simple two-fold anisotropy depends on $B_{\perp}^2$, the perpendicular component of the magnetic field with respect to the current direction. However, once the Fermi surface contains regions of varying curvature, additional harmonics arise due to the non-uniform orbital velocities and effective masses across different *k*-space directions [57,59]. In particular, segments with negative curvature generate secondary orbital loops in momentum space, which contribute to a higher anisotropic term to the AMR response [57]. At higher magnetic fields, transport becomes increasingly sensitive to the shape of the Fermi surface, and these effects become increasingly more pronounced as the field is increased.

In addition, the modulation of DOS near the Fermi energy has also been recently proposed in Kagome system such as $Co_3In_2S_2$ [60], in which the rotation of magnetic moment with external field leads to different band splitting that affects DOS near the Fermi energy, giving rise to anisotropic AMR. Furthermore, the symmetry of V 3d orbitals may be relevant to AMR anisotropy, as suggested in prior theoretical studies showing the influence of the orbital geometry on the angular dependence of scattering and transport under high magnetic fields [60,62,63]. These mechanisms involve a more complicated interplay between charge, spin, and orbital. To fully understand the four-fold anisotropy of the positive MR component in $GdV_6Sn_6$ and $NdV_6Sn_6$, experimental efforts in clarification of the Fermi surface and detailed theoretical studies are necessary.

**Conclusion:**

In summary, Nd substitution in $GdV_6Sn_6$ modifies the magnetic anisotropy without introducing significant structural distortion that changes lattice symmetry, offering a platform to investigate the coupling between transport, magnetism, and other exotic features of the Kagome lattice. The magnetotransport exhibits a crossover from negative to positive MR with increasing magnetic field, which is associated with magnetic scattering and orbital MR and features two- and four-fold anisotropy respectively. Substituting Nd for Gd switches the phase of the low-field two-fold anisotropy by tuning magnetic anisotropy, while preserving the phase of the high-field four-fold anisotropy. Our study of the previously unexplored anisotropic transport in non-magnetic V-based *R166* Kagome systems demonstrates how MR can be modulated by magnetic field and rare-earth substitution, opening a playground to probe, control, and design transport behaviours from the rich interplay of rich quantum properties in Kagome materials.

**Acknowledgment:**

This work was primarily supported by by the U.S. Department of Energy (DOE), Office of Science, Basic Energy Sciences program under Grant No. DE-SC0022006 (synthesis and transport measurements). We acknowledge the MonArk NSF Quantum Foundry for magnetic property measurements using MPMS3 SQUID, which is supported by the National Science Foundation (NSF) Q-AMASE-i program under NSF Award No. DMR-1906383. J. W. acknowledges the support from National Scientific Foundation under grant DMR-2316811 for single crystal XRD.

**Table 1.** EDS composition and structure parameters obtained from single crystal XRD.

| Nominal Composition | EDS Composition | Lattice parameters (Å) | | | Atoms | Atomic positions | | | Occ. |
|---|---|---|---|---|---|---|---|---|---|
| | | *a* | *b* | *c* | | *x* | *y* | *z* | |
| $GdV_6Sn_6$ | $GdV_{6.1}Sn_{5.8}$ | 5.519(1) | 5.519(1) | 9.170(1) | *R* | 0.0000 | 0.0000 | 0.0000 | 1 |
| | | | | | V | -0.5000 | 0.0000 | 0.2513 | 1 |
| | | | | | Sn1 | 0.0000 | 0.0000 | 0.1654 | 1 |
| | | | | | Sn2 | -0.3333 | 0.3333 | 0.0000 | 1 |
| | | | | | Sn3 | -0.3333 | 0.3333 | 0.5000 | 1 |
| $Nd_{0.05}Gd_{0.95}V_6Sn_6$ | $Nd_{0.05}Gd_{0.95}V_{6.4}Sn_{5.8}$ | 5.518(4) | 5.518(4) | 9.171(5) | *R* | 0.0000 | 0.0000 | 0.0000 | 1 |
| | | | | | V | -0.5000 | 0.0000 | 0.2513 | 1 |
| | | | | | Sn1 | 0.0000 | 0.0000 | 0.1654 | 1 |
| | | | | | Sn2 | -0.3333 | 0.3333 | 0.0000 | 1 |
| | | | | | Sn3 | -0.3333 | 0.3333 | 0.5000 | 1 |
| $Nd_{0.5}Gd_{0.5}V_6Sn_6$ | $Nd_{0.51}Gd_{0.49}V_{6.3}Sn_{6.1}$ | 5.539(4) | 5.539(4) | 9.215(1) | *R*1 | 1.0000 | 1.0000 | 0.5000 | 0.81 |
| | | | | | *R*2 | 1.0000 | 1.0000 | 0.6666 | 0.08 |
| | | | | | V | 0.5000 | 0.5000 | 0.7493 | 1 |
| | | | | | Sn1 | 1.0000 | 1.0000 | 1.0000 | 0.12 |
| | | | | | Sn2 | 0.3333 | 0.6666 | 1.0000 | 1 |
| | | | | | Sn3 | 0.6666 | 0.3333 | 0.5000 | 1 |
| | | | | | Sn4 | 1.0000 | 1.0000 | 0.8351 | 0.90 |
| $Nd_{0.75}Gd_{0.25}V_6Sn_6$ | $Nd_{0.25}Gd_{0.75}V_{6.9}Sn_{6.3}$ | 5.533(2) | 5.533(2) | 9.203(6) | *R*1 | 0.0000 | 1.0000 | 0.5000 | 0.82 |
| | | | | | *R*2 | 0.0000 | 1.0000 | 0.3396 | 0.12 |
| | | | | | V | 0.5000 | 1.0000 | 0.2496 | 1 |
| | | | | | Sn1 | 0.6666 | 1.3333 | 0.5000 | 1 |
| | | | | | Sn2 | 0.3333 | 0.6666 | 0.0000 | 1 |
| | | | | | Sn3 | 0.0000 | 1.0000 | 0.1667 | 1 |
| | | | | | Sn4 | 0.0000 | 1.0000 | 0.0000 | 0.26 |
| $NdV_6Sn_6$ | $NdV_{6.1}Sn_{6.3}$ | 5.546(6) | 5.546(6) | 9.214(8) | *R*1 | 0.0000 | 1.0000 | 0.5000 | 0.82 |
| | | | | | *R*2 | 0.0000 | 1.0000 | 0.6624 | 0.13 |
| | | | | | V | 0.5000 | 1.0000 | 0.7496 | 1 |
| | | | | | Sn1 | 0.6600 | 1.3333 | 1.0000 | 1 |
| | | | | | Sn2 | 0.0000 | 1.0000 | 1.0000 | 0.2 |
| | | | | | Sn3 | 0.3333 | 0.6666 | 0.5000 | 1 |
| | | | | | Sn4 | 0.0000 | 1.0000 | 0.8359 | 0.82 |

## References:

[1] C. S. Gruber and M. Abdel-Hafiez, Interplay of Electronic Orders in Topological Quantum Materials, ACS Mater. Au **5**, 72 (2025).

[2] G. R. Stewart, Unconventional superconductivity, Advances in Physics **66**, 75 (2017).

[3] M. M. Sharma, P. Sharma, N. K. Karn, and V. P. S. Awana, Comprehensive review on topological superconducting materials and interfaces, Supercond. Sci. Technol. **35**, 083003 (2022).

[4] Xianxin Wu, Tilman Schwemmer, Tobias Müller, Armando Consiglio, Giorgio Sangiovanni, Domenico Di Sante, Yasir Iqbal, Werner Hanke, Andreas P. Schnyder, M. Michael Denner, Mark H. Fischer, Titus Neupert, and Ronny Thomale, Nature of Unconventional Pairing in the Kagome Superconductors $AV_3Sb_5$ (A = K, Rb, Cs), Phys. Rev. Lett. **127**, 177001 (2021).

[5] S. C. Holbæk, M. H. Christensen, A. Kreisel, and B. M. Andersen, Unconventional superconductivity protected from disorder on the kagome lattice, Phys. Rev. B **108**, 144508 (2023).

[6] Y. Chen, J. Gaudet, G. G. Marcus, T. Nomoto, T. Chen, T. Tomita, M. Ikhlas, H. S. Suzuki, Y. Zhao, W. C. Chen, J. Strempfer8, R. Arita, S. Nakatsuji, and C. Broholm, Intertwined charge and spin density waves in a topological kagome material, Phys. Rev. Res. **6**, L032016 (2024).

[7] Seongyong Lee, Choongjae Won, Jimin Kim, Jonggyu Yoo, Sudong Park, Jonathan Denlinger, Chris Jozwiak, Aaron Bostwick, Eli Rotenberg, Nature of charge density wave in kagome metal $ScV_6Sn_6$, Npj Quantum Mater. **9**, 1 (2024).

[8] Masayoshi Fujihala, Katsuhiro Morita, Richard Mole, Setsuo Mitsuda, Takami Tohyama, Shin-ichiro Yano, Dehong Yu, Shigetoshi Sota, Tomohiko Kuwai, Akihiro Koda, Hirotaka Okabe, Hua Lee, Shinichi Itoh, Takafumi Hawai, Takatsugu Masuda, Hajime Sagayama, Akira Matsuo, Koichi Kindo, Seiko Ohira-Kawamura & Kenji Nakajima, Gapless spin liquid in a square-kagome lattice antiferromagnet, Nat Commun **11**, 3429 (2020).

[9] Y. Zhou, K. Kanoda, and T.-K. Ng, Quantum spin liquid states, Rev. Mod. Phys. **89**, 025003 (2017).

[10] C. Rüegg, A pinwheel without wind, Nature Phys **6**, 837 (2010).

[11] M. M. Sharma, Santosh Karki Chhetri, Gokul Acharya, David Graf, Dinesh Upreti, Sagar Dahal, Md Rafique Un Nabi, Sumaya Rahman, Josh Sakon, Hugh O.H. Churchill, Jin Hu, Quantum oscillation studies of the nodal line semimetal $Ni_3In_2S_{2-x}Se_x$, Acta Materialia **289**, 120884 (2025).

[12] Hongwei Fang, Meng Lyu, Hao Su, Jian Yuan, Yiwei Li, Lixuan Xu, Shuai Liu, Liyang Wei, Xinqi Liu, Haifeng Yang, Qi Yao, Meixiao Wang, Yanfeng Guo, Wujun Shi, Yulin Chen, Enke Liu and Zhongkai Liu , Record-high mobility and extreme magnetoresistance on kagome-lattice in compensated semimetal $Ni_3In_2S_2$, Sci. China Mater. **66**, 2032 (2023).

[13] Man Li, Qi Wang, Guangwei Wang, Zhihong Yuan, Wenhua Song, Rui Lou, Zhengtai Liu,Yaobo Huang, Zhonghao Liu, Hechang Lei, Zhiping Yin and Shancai Wang, Dirac cone, flat band and saddle point in kagome magnet $YMn_6Sn_6$, Nat Commun **12**, 3129 (2021).

[14] Zhiyong Lin, Chongze Wang, Pengdong Wang, Seho Yi, Lin Li, Qiang Zhang, Yifan Wang, Zhongyi Wang, Hao Huang, Yan Sun, Yaobo Huang, Dawei Shen, Donglai Feng, Zhe Sun, Jun-Hyung Cho, Changgan Zeng, and Zhenyu Zhang, Dirac fermions in antiferromagnetic FeSn kagome lattices with combined space inversion and time-reversal symmetry, Phys. Rev. B **102**, 155103 (2020).

[15] M. Li, H. Ma, R. Lou, and S. Wang, Electronic band structures of topological kagome materials, Chinese Phys. B **34**, 017101 (2025).

[16] B. He, M. Yao, Y. Pan, K. E. Arpino, D. Chen, F. M. Serrano-Sanchez, S. Ju, M. Shi, Y. Sun, and C. Felser, Enhanced Weyl semimetal signature in $Co_3Sn_2S_2$ Kagome ferromagnet by chlorine doping, Commun Mater **5**, 275 (2024).

[17] D. F. Liu, A. J. Liang, E. K. Liu, Q. N. Xu, Y. W. Li , C. Chen, D. Pei,W. J. Shi, S. K. Mo, P. Dudin, T. Kim, C. Cacho, G. Li, Y. Sun, L. X. Yang, Z. K. Liu, S. S. P. Parkin, C. Felser, and Y. L. Chen, Magnetic Weyl semimetal phase in a Kagomé crystal, Science **365**, 1282 (2019).

[18] S. Kumar Pradhan, S. Pradhan, P. Mal, P. Rambabu, A. Lakhani, B. Das, B. Lingam Chittari, G. R. Turpu, and P. Das, Endless Dirac nodal lines and high mobility in kagome semimetal $Ni_3In_2Se_2$ : a theoretical and experimental study, J. Phys.: Condens. Matter **36**, 445601 (2024).

[19] Yong Hu, Xianxin Wu, Brenden R. Ortiz, Sailong Ju, Xinloong Han, Junzhang Ma, Nicholas C. Plumb, Milan Radovic, Ronny Thomale, Stephen D. Wilson, Andreas P. Schnyder, and Ming Shi, Rich nature of Van Hove singularities in Kagome superconductor $CsV_3Sb_5$, Nat Commun **13**, 2220 (2022).

[20] Yang Luo, Yulei Han, Jinjin Liu, Hui Chen, Zihao Huang, Linwei Huai, Hongyu Li, Bingqian Wang, Jianchang Shen, Shuhan Ding, Zeyu Li, Shuting Peng, Zhiyuan Wei, Yu Miao, Xiupeng Sun, Zhipeng Ou, Ziji Xiang, Makoto Hashimoto, Donghui Lu, Yugui Yao, Haitao Yang, Xianhui Chen, Hong-Jun Gao, Zhenhua Qiao, Zhiwei Wang, and Junfeng He, A unique van Hove singularity in kagome superconductor $CsV_{3-x}Ta_xSb_5$ with enhanced superconductivity, Nat Commun **14**, 3819 (2023).

[21] Mingu Kang, Shiang Fang, Linda Ye, Hoi Chun Po, Jonathan Denlinger, Chris Jozwiak, Aaron Bostwick, Eli Rotenberg, Efthimios Kaxiras, Joseph G. Checkelsky and Riccardo Comin, Topological flat bands in frustrated kagome lattice CoSn, Nat Commun **11**, 4004 (2020).

[22] S. Okamoto, N. Mohanta, E. Dagotto, and D. N. Sheng, Topological flat bands in a kagome lattice multiorbital system, Commun Phys **5**, 1 (2022).

[23] C.-X. Liu, S.-C. Zhang, and X.-L. Qi, The Quantum Anomalous Hall Effect: Theory and Experiment, Annual Review of Condensed Matter Physics 7, 301 (2016).

[24] Cui-Zu Chang, Jinsong Zhang, Xiao Feng, Jie Shen, Zuocheng Zhang, Minghua Guo, Kang Li, Yunbo Ou, Pang Wei, Li-Li Wang, Zhong-Qing Ji, Yang Feng, Shuaihua Ji, Xi Chen, Jinfeng Jia, Xi Dai, Zhong Fang, Shou-Cheng Zhang, Ke He, Yayu Wang, Li Lu, Xu-Cun Ma, and Qi-Kun Xue, Experimental Observation of the Quantum Anomalous Hall Effect in a Magnetic Topological Insulator, Science **340**, 167 (2013).

[25] Y. Wan, J. Li, and Q. Liu, Topological magnetoelectric response in ferromagnetic axion insulators, National Science Review **11**, nwac138 (2024).

[26] J. Wang, B. Lian, X.-L. Qi, and S.-C. Zhang, Quantized topological magnetoelectric effect of the zero-plateau quantum anomalous Hall state, Phys. Rev. B **92**, 081107 (2015).

[27] B. C. Sales, W. R. Meier, A. F. May, J. Xing, J.-Q. Yan, S. Gao, Y. H. Liu, M. B. Stone, A. D. Christianson, Q. Zhang and M. A. McGuire, Tuning the flat bands of the kagome metal CoSn with Fe, In, or Ni doping, Phys. Rev. Mater. **5**, 044202 (2021).

[28] Q. Wang, Y. Xu, R. Lou, Z. Liu, M. Li, Y. Huang, D. Shen, H. Weng, S. Wang, and H. Lei, Large intrinsic anomalous Hall effect in half-metallic ferromagnet $Co_3Sn_2S_2$ with magnetic Weyl fermions, Nat Commun **9**, 3681 (2018).

[29] K. Kumar, M. M. Sharma, and V. P. S. Awana, Weak antilocalization and ferromagnetism in magnetic Weyl semimetal $Co_3Sn_2S_2$, Journal of Applied Physics **133**, 023901 (2023).

[30] Jia-Xin Yin, Songtian S. Zhang, Guoqing Chang, Qi Wang, Stepan S. Tsirkin, Zurab Guguchia, Biao Lian, Huibin Zhou, Kun Jiang, Ilya Belopolski, Nana Shumiya, Daniel Multer, Maksim Litskevich, Tyler A. Cochran, Hsin Lin, Ziqiang Wang, Titus Neupert, Shuang Jia, Hechang Lei and M. Zahid Hasan, Negative flat band magnetism in a spin–orbit-coupled correlated kagome magnet, Nat. Phys. **15**, 443 (2019).

[31] W. Ma, X. Xu, Z. Wang, H. Zhou, M. Marshall, Z. Qu, W. Xie, and S. Jia, Anomalous Hall effect in the distorted kagome magnets (Nd,Sm)$Mn_6Sn_6$, Phys. Rev. B **103**, 235109 (2021).

[32] X. Xu, J.-X. Yin, Z. Qu, and S. Jia, Quantum interactions in topological R166 kagome magnet, Rep. Prog. Phys. **86**, 114502 (2023).

[33] G. Venturini, R. Welter, B. Malaman, and E. Ressouche, Magnetic structure of YMn6Ge6 and room temperature magnetic structure of $LuMn_6Sn_6$ obtained from neutron diffraction study, Journal of Alloys and Compounds **200**, 51 (1993).

[34] G. Venturini, B. C. E. Idrissi, and B. Malaman, Magnetic properties of $RMn_6Sn_6$ (R = Sc, Y, Gd−Tm, Lu) compounds with $HfFe_6Ge_6$ type structure, Journal of Magnetism and Magnetic Materials **94**, 35 (1991).

[35] K. Fruhling, A. Streeter, S. Mardanya, X. Wang, P. Baral, O. Zaharko, I. I. Mazin, S. Chowdhury, W. D. Ratcliff, and F. Tafti, Topological Hall effect induced by chiral fluctuations in $ErMn_6Sn_6$, Phys. Rev. Materials **8**, 094411 (2024).

[36] Wenlong Ma, Xitong Xu, Jia-Xin Yin, Hui Yang, Huibin Zhou, Zi-Jia Cheng, Yuqing Huang, Zhe Qu, Fa Wang, M. Zahid Hasan and Shuang Jia, Rare Earth Engineering in $RMn_6Sn_6$ ( R = Gd−Tm, Lu) Topological Kagome Magnets, Phys. Rev. Lett. **126**, 246602 (2021).

[37] Jia-Xin Yin, Wenlong Ma, Tyler A Cochran, Xitong Xu, Songtian S Zhang, Hung-Ju Tien, Nana Shumiya, Guangming Cheng, Kun Jiang, Biao Lian, Zhida Song, Guoqing Chang, Ilya Belopolski, Daniel Multer, Maksim Litskevich, Zi-Jia Cheng, Xian P Yang, Bianca Swidler, Huibin Zhou, Hsin Lin, Titus Neupert, Ziqiang Wang, Nan Yao, Tay-Rong Chang, Shuang Jia, M Zahid Hasan, Quantum-limit Chern topological magnetism in $TbMn_6Sn_6$, Nature **583**, 533 (2020).

[38] Yishui Zhou, Min-Kai Lee, Sabreen Hammouda, Sheetal Devi, Shin-IchiroYano, Romain Sibille, Oksana Zaharko, Wolfgang Schmidt, Karin Schmalzl, Ketty Beauvois, Eric Ressouche, Po-Chun Chang, Chun-Hao Huang, Lieh-Jeng Chang, Thomas Brückel, and Yixi Su, Ground-state magnetic structures of topological kagome metals $RV_6Sn_6$ ( R = Tb , Dy , Ho , Er ), Phys. Rev. Research **6**, 043291 (2024).

[39] Y. Hu, X. Wu, Y. Yang, S. Gao, N. C. Plumb, A. P. Schnyder, W. Xie, J. Ma, and M. Shi, Tunable topological Dirac surface states and van Hove singularities in kagome metal $GdV_6Sn_6$, Sci. Adv. **8**, eadd2024 (2022).

[40] Domenico Di Sante, Chiara Bigi, Philipp Eck, Stefan Enzner, Armando Consiglio, Ganesh Pokharel, Pietro Carrara, Pasquale Orgiani, Vincent Polewczyk, Jun Fujii, Phil D. C.

King, Ivana Vobornik, Giorgio Rossi, Ilija Zeljkovic, Stephen D. Wilson, Ronny Thomale, Giorgio Sangiovanni, Giancarlo Panaccione and Federico Mazzola, Flat band separation and robust spin Berry curvature in bilayer kagome metals, Nat. Phys. **19**, 1135 (2023).

[41] K. Guo, J. Ye, S. Guan, and S. Jia, Triangular Kondo lattice in $YbV_6Sn_6$ and its quantum critical behavior in a magnetic field, Phys. Rev. B **107**, 205151 (2023).

[42] Xing Huang, Zhiqiang Cui, Chaoxin Huang, Mengwu Huo, Hui Liu, Jingyuan Li, Feixiang Liang, Lan Chen, Hualei Sun, Bing Shen, Yunwei Zhang, and Meng Wang, Anisotropic magnetism and electronic properties of the kagome metal $SmV_6Sn_6$, Phys. Rev. Materials **7**, 054403 (2023).

[43] C. Dhital, G. Pokharel, B. Wilson, I. Kendrick, M. M. Asmar, D. Graf, J. Guerrero-Sanchez, R. Gonzalez-Hernandez, and S. D. Wilson, Fermi surface of the magnetic kagome compound $GdV_6Sn_6$ investigated using de Haas–van Alphen oscillations, Phys. Rev. B **109**, 235145 (2024).

[44] Z. Porter, G. Pokharel, J.-W. Kim, P. J. Ryan, and S. D. Wilson, Incommensurate magnetic order in the Z2 kagome metal $GdV_6Sn_6$, Phys. Rev. B **108**, 035134 (2023).

[45] Gokul Acharya, Bimal Neupane, Chia-Hsiu Hsu, Xian P. Yang, David Graf, Eun Sang Choi, Krishna Pandey, Md Rafique Un Nabi, Santosh Karki Chhetri, Rabindra Basnet, Sumaya Rahman, Jian Wang, Zhengxin Hu, Bo Da, Hugh O. H Churchill, Guoqing Chang, M. Zahid Hasan, Yuanxi Wang, and Jin Hu, Insulator-to-Metal Transition and Isotropic Gigantic Magnetoresistance in Layered Magnetic Semiconductors, Advanced Materials **36**, 2410655 (2024).

[46] Simon X. M. Riberolles, Tianxiong Han, Tyler J. Slade, J. M. Wilde, A. Sapkota, Wei Tian, Qiang Zhang, D. L. Abernathy, L. D. Sanjeewa, S. L. Bud'ko, P. C. Canfield, R. J. McQueeney, and B. G. Ueland, New insight into tuning magnetic phases of $RMn_6Sn_6$ kagome metals, Npj Quantum Mater. **9**, 42 (2024).

[47] See Supplemental Material at [URL will be inserted by publisher] for additional data of $Nd_xGd_{1-x}V_6Sn_6$ crystals.

[48] H. Ishikawa, T. Yajima, M. Kawamura, H. Mitamura, and K. Kindo, $GdV_6Sn_6$: A Multi-carrier Metal with Non-magnetic 3d-electron Kagome Bands and 4*f*-electron Magnetism, J. Phys. Soc. Jpn. **90**, 124704 (2021).

[49] G. Pokharel, S. M. L. Teicher, B. R. Ortiz, P. M. Sarte, G. Wu, S. Peng, J. He, R. Seshadri, and S. D. Wilson, Electronic properties of the topological kagome metals $YV_6Sn_6$ and $GdV_6Sn_6$, Phys. Rev. B **104**, 235139 (2021).

[50] Santosh Karki Chhetri, Gokul Acharya, David Graf, Rabindra Basnet, Sumaya Rahman, M. M. Sharma, Dinesh Upreti, Md Rafique Un Nabi, Serhii Kryvyi, Josh Sakon, Mansour Mortazavi, Bo Da, Hugh Churchill, and Jin Hu, Large negative magnetoresistance in antiferromagnetic $Gd_2Se_3$, Phys. Rev. B **111**, 014431 (2025).

[51] J. Hu, S.-Y. Xu, N. Ni, and Z. Mao, Transport of Topological Semimetals, Annual Review of Materials Research **49**, 207 (2019).

[52] Kaustuv Manna, Lukas Muechler, Ting-Hui Kao, Rolf Stinshoff, Yang Zhang, Johannes Gooth, Nitesh Kumar, Guido Kreiner, Klaus Koepernik, Roberto Car, Jürgen Kübler, Gerhard H. Fecher, Chandra Shekhar, Yan Sun, and Claudia Felser, From Colossal to Zero: Controlling

the Anomalous Hall Effect in Magnetic Heusler Compounds via Berry Curvature Design, Phys. Rev. X **8**, 041045 (2018).

[53] P. Li, C. Jin, E. Y. Jiang, and H. L. Bai, Origin of the twofold and fourfold symmetric anisotropic magnetoresistance in epitaxial $Fe_3O_4$ films, Journal of Applied Physics **108**, 093921 (2010).

[54] J. Li, S. L. Li, Z. W. Wu, S. Li, H. F. Chu, J. Wang, Y. Zhang, H. Y. Tian, and D. N. Zheng, A phenomenological approach to the anisotropic magnetoresistance and planar Hall effect in tetragonal $La_{2/3}Ca_{1/3}MnO_3$ thin films, J. Phys.: Condens. Matter **22**, 146006 (2010).

[55] S. Ghosh, A. Low, N. Devaraj, S. Changdar, A. Narayan, and S. Thirupathaiah, Extremely large and angle-dependent magnetoresistance in kagome Dirac semimetal $RFe_6Sn_6$ (R=Ho, Dy), Journal of Alloys and Compounds **1040**, 183506 (2025).

[56] M. N. Ali, L. M. Schoop, C. Garg, J. M. Lippmann, E. Lara, B. Lotsch, and S. S. P. Parkin, Butterfly magnetoresistance, quasi-2D Dirac Fermi surface and topological phase transition in ZrSiS, Science Advances **2**, e1601742 (2016).

[57] N. P. Ong, Geometric interpretation of the weak-field Hall conductivity in two-dimensional metals with arbitrary Fermi surface, Phys. Rev. B **43**, 193 (1991).

[58] Jialu Wang, Haiyang Yang, Linchao Ding, Wei You, Chuanying Xi, Jie Cheng, Zhixiang Shi, Chao Cao, Yongkang Luo, Zengwei Zhu, Jianhui Dai, Mingliang Tian, Yuke Li, Angle-dependent magnetoresistance and its implications for Lifshitz transition in $W_2As_3$, Npj Quantum Mater. **4**, 58 (2019).

[59] S. Zhang, Q. Wu, Y. Liu, and O. V. Yazyev, Magnetoresistance from Fermi surface topology, Phys. Rev. B **99**, 035142 (2019).

[60] Senhao Lv, Hui Guo, Qi Qi, Yuhui Li, Guojing Hu, Qi Zheng, Ruwen Wang, Nan Si, Ke Zhu, Zhen Zhao, Yechao Han, Weiqi Yu, Guoyu Xian, Li Huang, Lihong Bao, Xiao Lin, Jinbo Pan, Shixuan Du, Jun He, Haitao Yang, and Hong-Jun Gao, Field-Induced Butterfly-Like Anisotropic Magnetoresistance in a Kagome Semimetal $Co_3In_2S_2$, Adv Funct Materials **35**, 2412876 (2025).

[61] Y. Dai, Y. W. Zhao, L. Ma, M. Tang, X. P. Qiu, Y. Liu, Z. Yuan, and S. M. Zhou, Fourfold Anisotropic Magnetoresistance of $L1_0$FePt Due to Relaxation Time Anisotropy, Phys. Rev. Lett. **128**, 247202 (2022).

[62] H.-W. Ko, H.-J. Park, G. Go, J. H. Oh, K.-W. Kim, and K.-J. Lee, Role of orbital hybridization in anisotropic magnetoresistance, Phys. Rev. B **101**, 184413 (2020).

[63] D. Li, F. Pauly, and A. Smogunov, Giant anisotropic magnetoresistance through a tilted molecular π -orbital, Phys. Rev. Research **2**, 033184 (2020).

**Figures:**

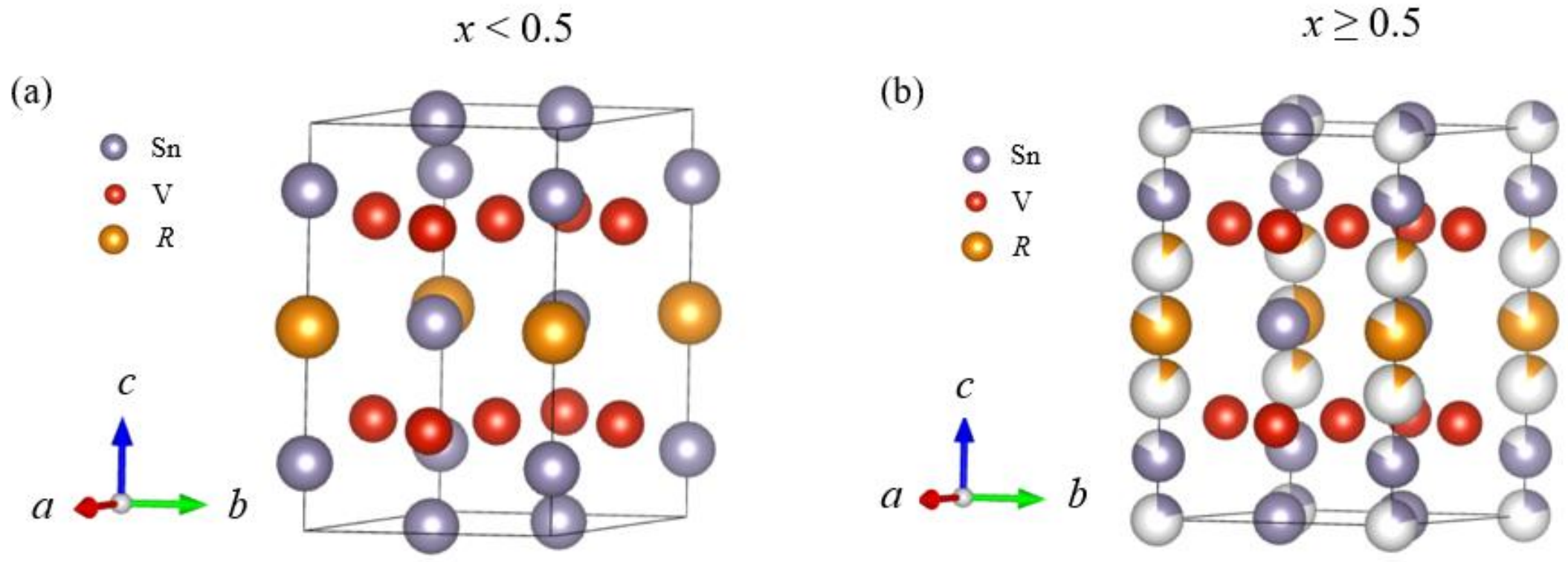

**Fig. 1:** Unit cell structure of the synthesized $Nd_xGd_{1-x}V_6Sn_6$ single crystals (a) unit cell structure for x < 0.5, where the atomic sites of the rare earth elements are not split (b) unit cell structure for x ≥ 0.5, where the rare earth elements partially occupy two different atomic sites.

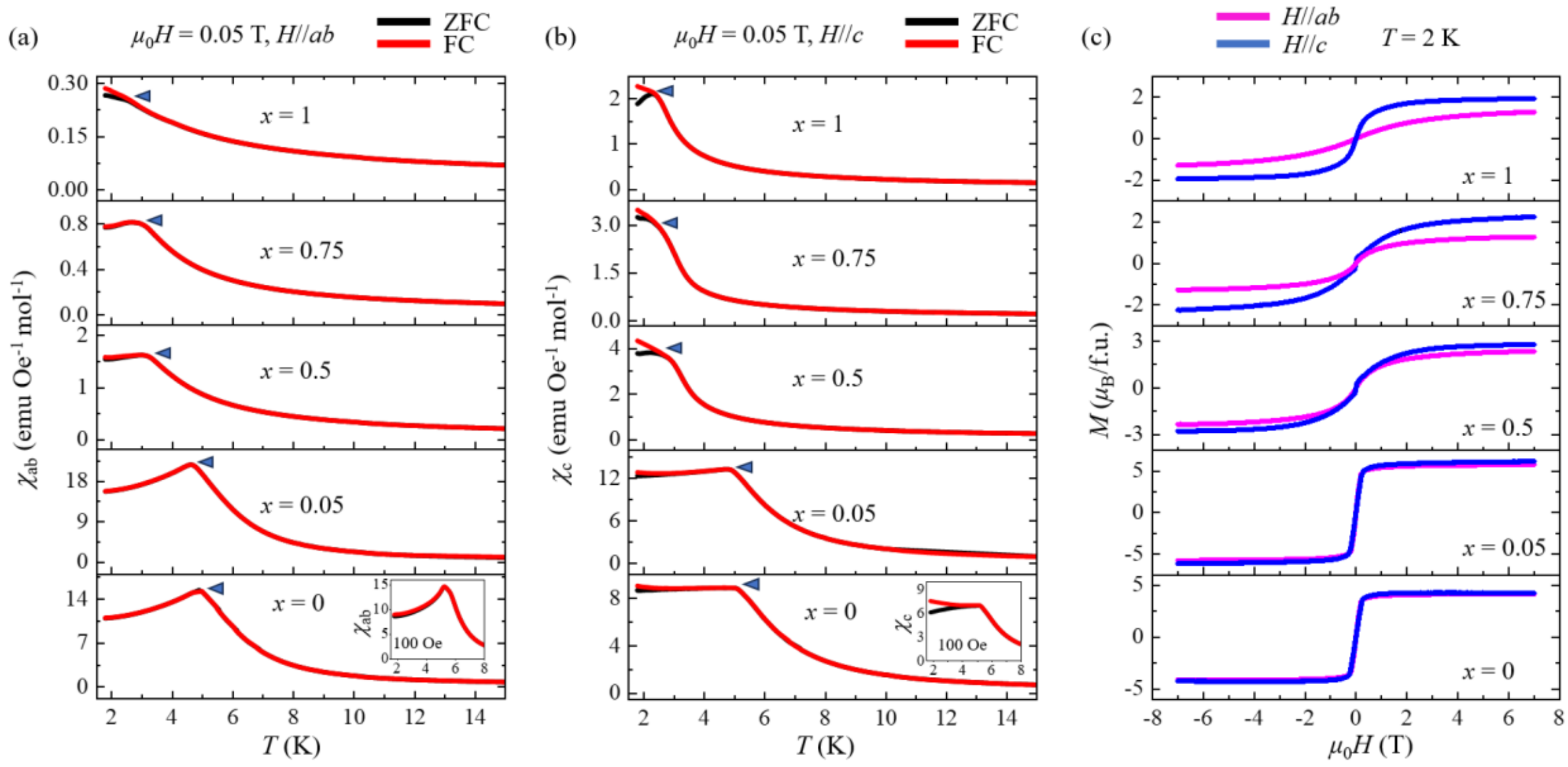

**Fig. 2:** Temperature-dependent magnetic susceptibility ($\chi$-*T*) of $Nd_xGd_{1-x}V_6Sn_6$ (x = 0, 0.05, 0.5, 0.75, and 1) single crystals measured under *ZFC* and *FC* conditions in an applied magnetic field of 500 Oe with (a) *H* || *ab* plane and (b) *H* || *c*-axis. The insets in the lower panels show the magnified view of the $\chi$-*T* curves near the magnetic transition temperature measured at 100 Oe. (c) Isothermal magnetization ($M$-$\mu_0H$) curves recorded at 2 K for both field orientations (*H* || *ab* and H || *c*).

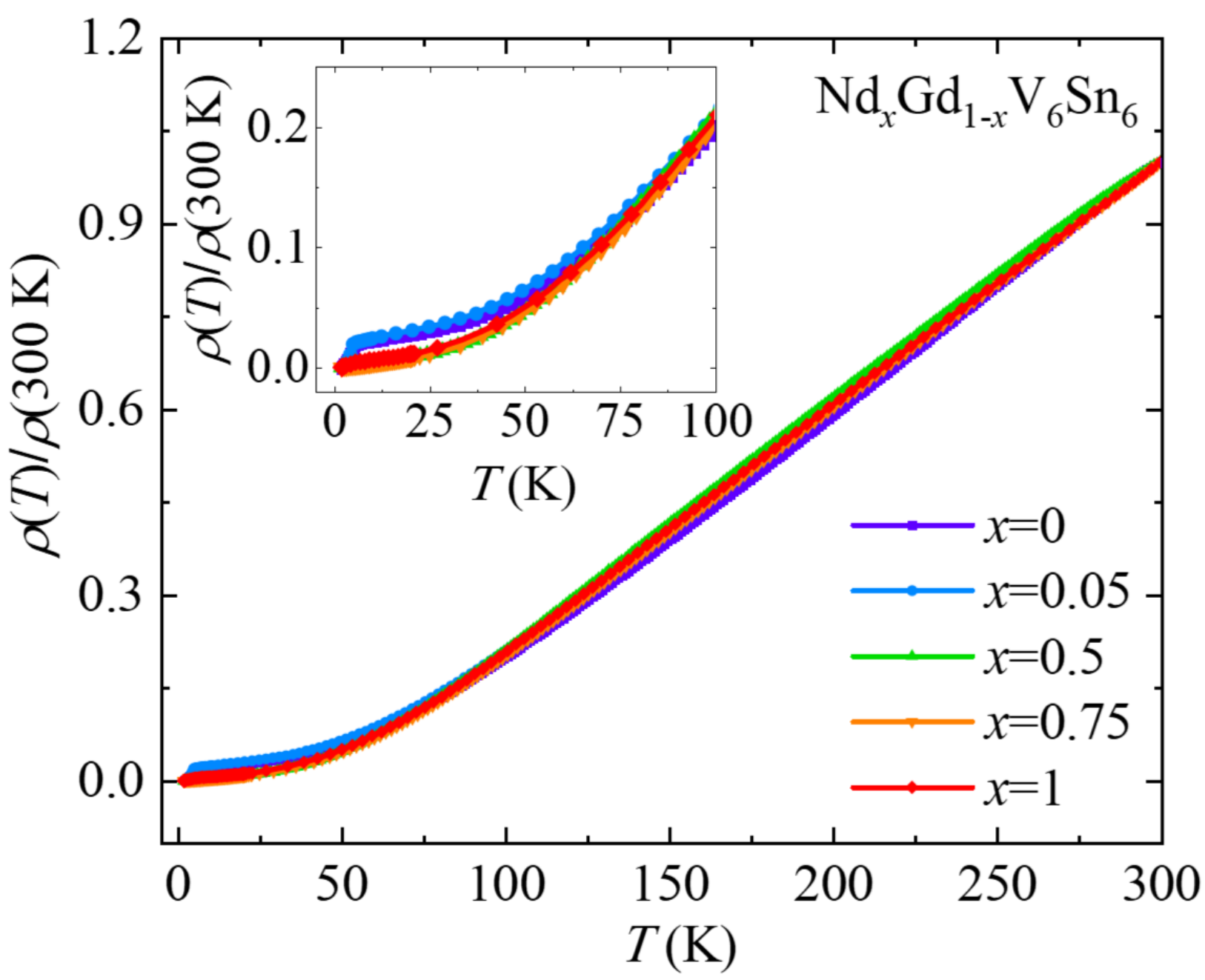

**Fig. 3:** Temperature-dependent normalized resistivity (ρ(T)/ρ(300)) plot for $Nd_xGd_{1-x}V_6Sn_6$ (x = 0, 0.05, 0.5, 0.75, and 1) single crystals. The inset shows a zoomed view of normalized resistivity to show the difference in the resistivity curve near the ordering temperature.

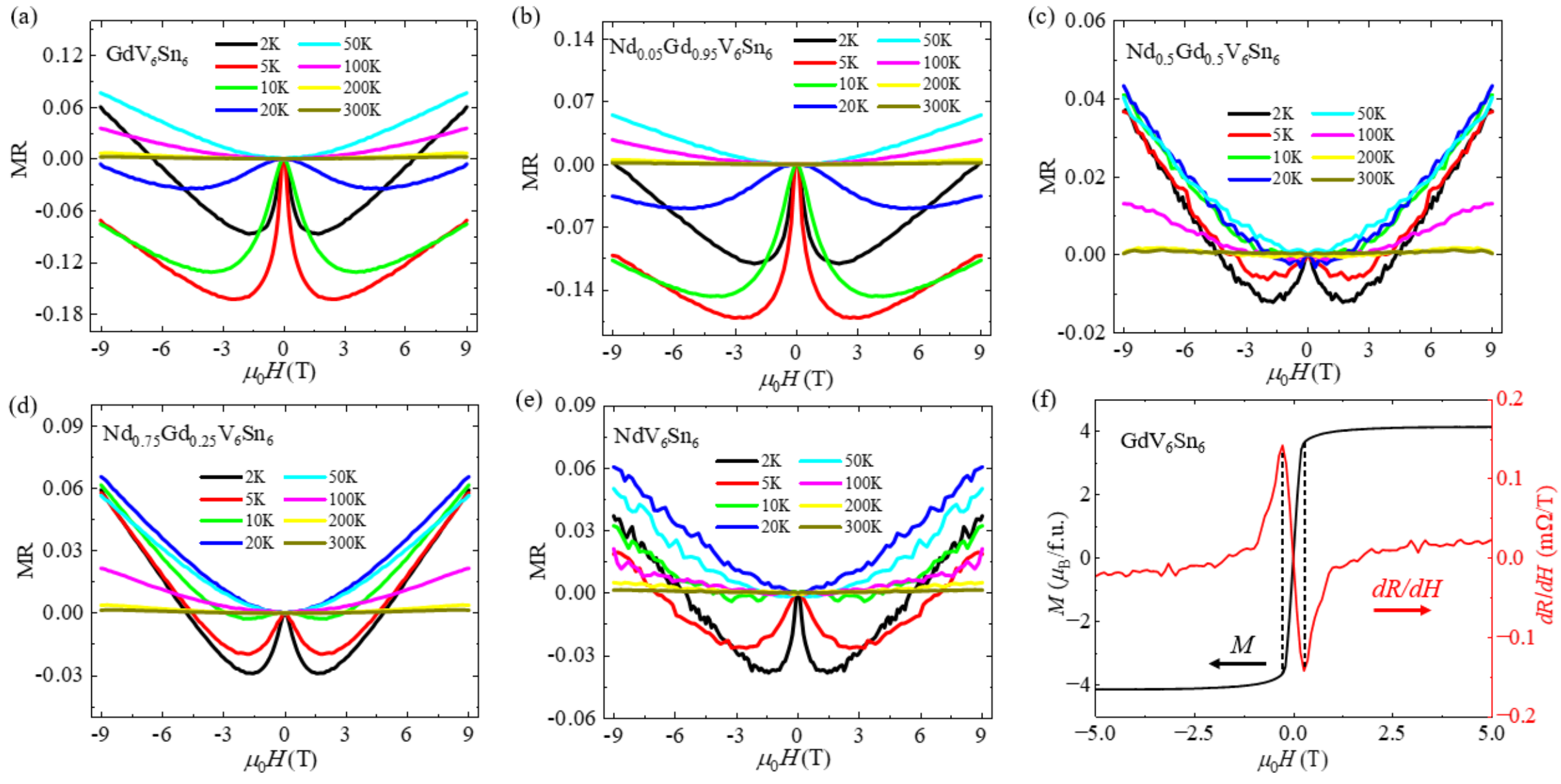

**Fig. 4:** MR as a function of applied magnetic field ($\mu_0 H$) measured at various temperatures ranging from 2 K to 300 K for (a) $GdV_6Sn_6$, (b) $Nd_{0.05}Gd_{0.95}V_6Sn_6$, (c) $Nd_{0.5}Gd_{0.5}V_6Sn_6$, (d) $Nd_{0.75}Gd_{0.25}V_6Sn_6$, and (e) $NdV_6Sn_6$. (f) Comparison between the field derivative of resistivity ($d\rho/dH$) and magnetization ($M$) for $GdV_6Sn_6$, illustrating the magnetic origin of the observed negative MR.

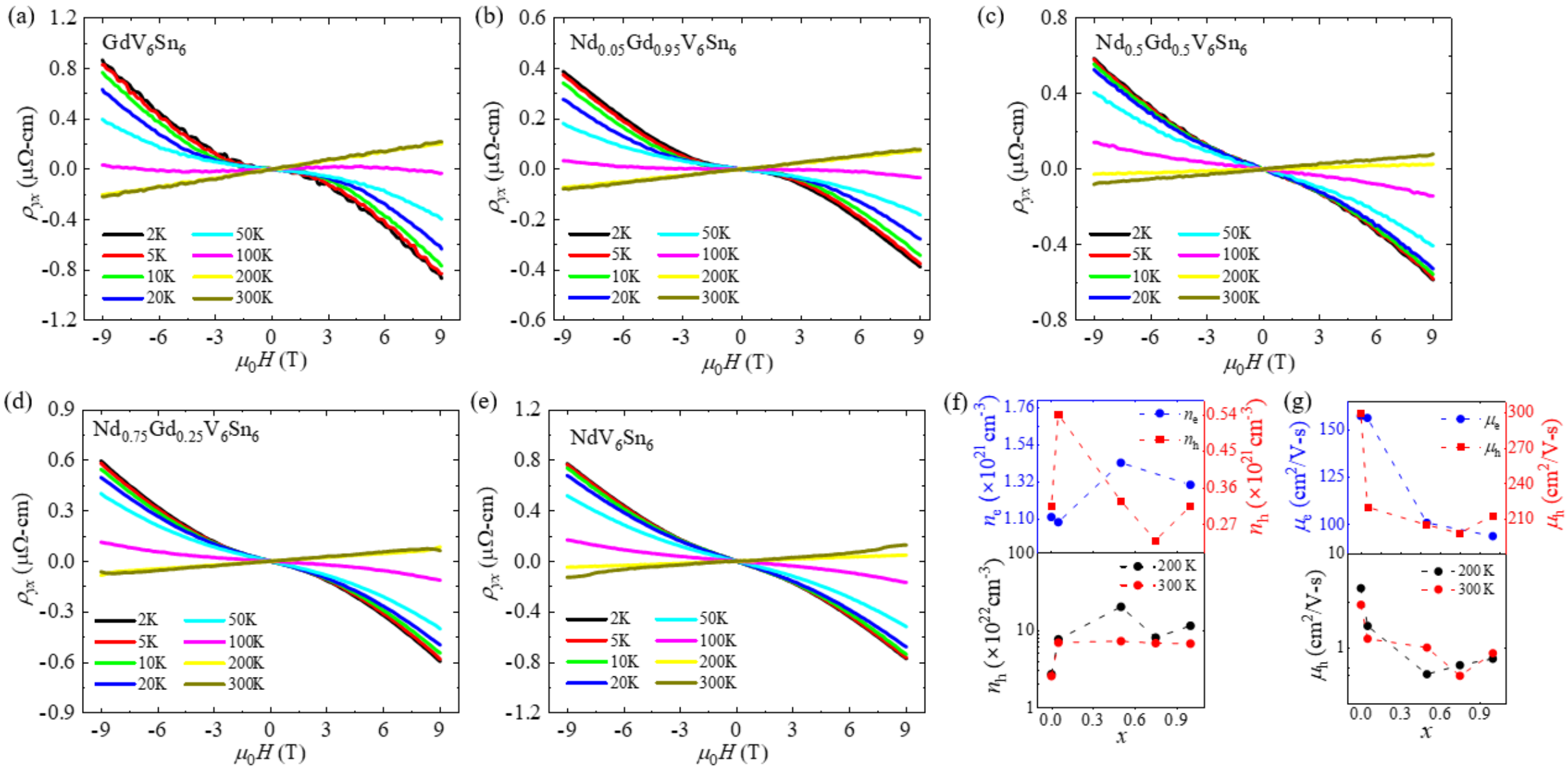

**Fig. 5:** Hall resistivity ($\rho_{yx}$) as a function of applied magnetic field ($\mu_0 H$) measured at various temperatures ranging from 2 K to 300 K for (a) $GdV_6Sn_6$, (b) $Nd_{0.05}Gd_{0.95}V_6Sn_6$, (c) $Nd_{0.5}Gd_{0.5}V_6Sn_6$, (d) $Nd_{0.75}Gd_{0.25}V_6Sn_6$, and (e) $NdV_6Sn_6$. (f) The upper panel shows electron and hole concentration at 100 K for different Nd content, the lower panel shows the hole concentration with respect to Nd content at 200 K and 300 K. (g) The upper panel shows electron and hole mobility at 100 K for different Nd content, the lower panel shows the hole mobility with respect to Nd content at 200 K and 300 K.

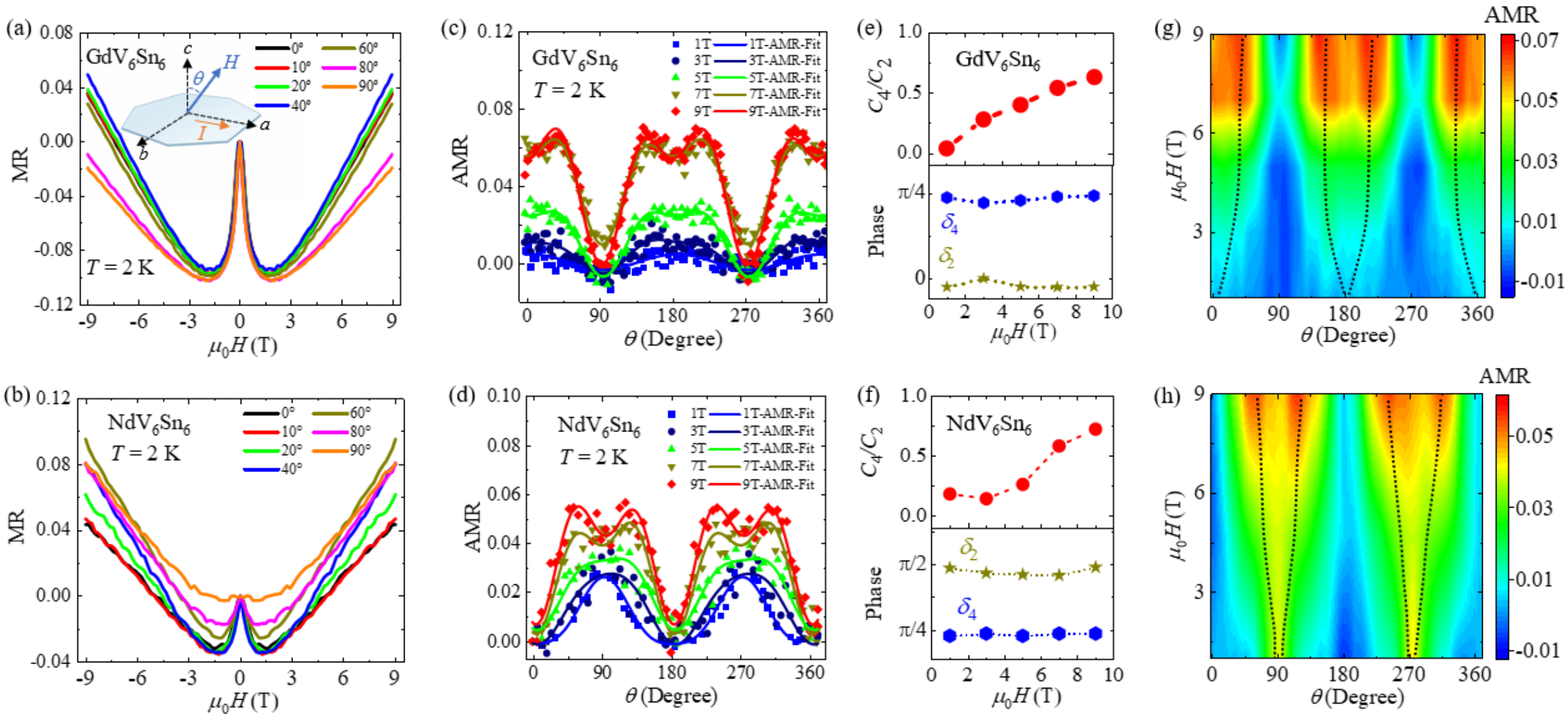

**Fig. 6:** MR as a function of applied magnetic field ($\mu_0 H$) at 2 K measured at various magnetic field orientations with respect to applied current (a) $GdV_6Sn_6$, the inset shows a schematic of magnetic field rotation (b) $Nd_{0.05}Gd_{0.95}V_6Sn_6$. (c) & (d) angle-dependent AMR measurement at 2 K under various magnetic fields for 1 T to 9 T for $GdV_6Sn_6$ and $NdV_6Sn_6$. (e) & (f) Upper panel is showing the ratio of the coefficient associated with fourfold anisotropy at different magnetic fields, while the lower panel is showing the field dependence phase factors associated with twofold anisotropy and fourfold anisotropy for $GdV_6Sn_6$ and $NdV_6Sn_6$. (g) & (h) Contour plot of AMR where the black dotted line shows the evolution of the fourfold anisotropy as the magnetic field increased in $GdV_6Sn_6$.